\documentclass[twocolumn,aps,prb]{revtex4}
\usepackage{amsmath,amssymb}
\usepackage{graphicx}
\usepackage{epstopdf}
\usepackage[usenames]{color}
\usepackage[]{natbib}

\newcommand{\be}{\begin{equation}}
\newcommand{\ee}{\end{equation}}
\newcommand{\bk}{{{\bf{k}}}}

\newcommand{\bea}{\begin{eqnarray}}
\newcommand{\eea}{\end{eqnarray}}
\newcommand{\ra}{\rangle}
\newcommand{\la}{\langle}

\newcommand{\upa}{\uparrow}
\newcommand{\dna}{\downarrow}

\newcommand{\dg}{{\dagger}}
\newcommand{\pdg}{{\phantom\dagger}}

\begin{document}

\title{Theory of metallic double perovskites with spin orbit coupling and strong correlations; application to ferrimagnetic Ba$_2$FeReO$_6$}
\author{Ashley Cook$^1$}
\author{Arun Paramekanti$^{1,2}$}
\affiliation{$^1$Department of Physics, University of Toronto, Toronto, Ontario, Canada M5S 1A7}
\affiliation{$^2$Canadian Institute for Advanced Research, Toronto, Ontario, M5G 1Z8, Canada}
\begin{abstract}
We consider a model of the double perovskite Ba$_2$FeReO$_6$, a room temperature ferrimagnet
with correlated and spin-orbit coupled Re t$_{2g}$ electrons moving in the
background of Fe moments stabilized by Hund's coupling. 
We show that for such 3d/5d double perovskites, strong correlations on the 5d-element (Re) are essential in driving a 
half-metallic ground state. Incorporating both strong
spin-orbit coupling and the Hubbard
repulsion on Re leads to a band structure consistent with ab initio calculations. Using our model, we find a large
spin polarization at the Fermi level, and obtain a semi-quantitative understanding
of the saturation magnetization of Ba$_2$FeReO$_6$, as well as X-ray magnetic circular dichroism data indicating a 
significant orbital magnetization.
Based on the orbital populations obtained in our theory, we predict a specific doping dependence to the tetragonal distortion
accompanying ferrimagnetic order.
Finally, the combination
of a net magnetization and spin-orbit interactions is shown to induce Weyl nodes in the band structure, and we
predict a significant intrinsic anomalous Hall effect in hole-doped Ba$_2$FeReO$_6$. The uncovered interplay of
strong correlations and spin-orbit coupling lends partial support to our previous work, which used a local moment
description to capture the spin wave dispersion found in neutron scattering measurements. Our work
is of interest in the broader context of understanding metallic double 
perovskites which are of fundamental importance and of possible relevance to spintronic applications.
 \end{abstract}
\maketitle

\section{Introduction} Double perovskite (DP) materials A$_2$BB'O$_6$, where the transition metal ions B and B' reside 
on the two sublattices of a cubic lattice, can realize many complex phases.\cite{Serrate_Review_JPCM2007} Metallic
variants, such as Sr$_2$FeMoO$_6$, provide us with the simplest multi-orbital examples of ferrimagnetic order \cite{Kobayashi1998}
kinetically stabilized by the Pauli exclusion 
principle.\cite{Sarma_PRL2000,Sarma_PRB2001,Jackeli2003,Chattopadhyay_PRB2001,Phillips_PRB2003,Brey2006,Majumdar2009,Erten_SFMO_PRL2011,Das_PRB2011} 
Insulating variants where only the
B'-site ion is magnetic, such as Ba$_2$YMoO$_6$ and La$_2$LiMoO$_6$, provide material examples of quantum mechanical moments living on the
geometrically frustrated face-centered cubic lattice.\cite{Aharen_PRB2010,BYMO_Neutron_PRB2011,Chen2010,DoddsPRB2011,Chen2011}
Metallic DPs, such as Sr$_2$FeMoO$_6$, are also of significant 
technological importance, being room temperature ferrimagnets with half-metallic band structures and a large spin 
polarization which is useful for spintronic applications.\cite{Zutic2004,Alff_Springer2007}
Metallic 3d/5d
DPs are of particular interest in this regard since they appear to have strongly reduced
B/B' site mixing; samples of Ba$_2$FeReO$_6$ studied in previous work \cite{Plumb_PRB2013} have low
$< 1\%$ anti-site disorder. Such anti-site disorder, which is common in other DPs and which is detrimental to spintronic
applications, appears to be alleviated in 3d/5d DPs by the B/B' ionic size mismatch suggesting that they might be better
suited for applications. However, such 
3d/5d DPs require us to confront the twin aspects of strong correlations and strong spin-orbit coupling,
topics at the forefront of fundamental research \cite{Iridate_Review} motivated by the possibility of stabilizing
states such as fractionalized topological insulators (TIs),\cite{Levin_PRL2009,Pesin_NPhys2010,Maciejko_PRL2010,Swingle_PRB2011} or 
Weyl semimetals.\cite{Wan_PRB2011,Burkov_PRL2011,YRan_PRB2011,Witczak2011}

\begin{figure}[t]
    \includegraphics[scale=0.5]{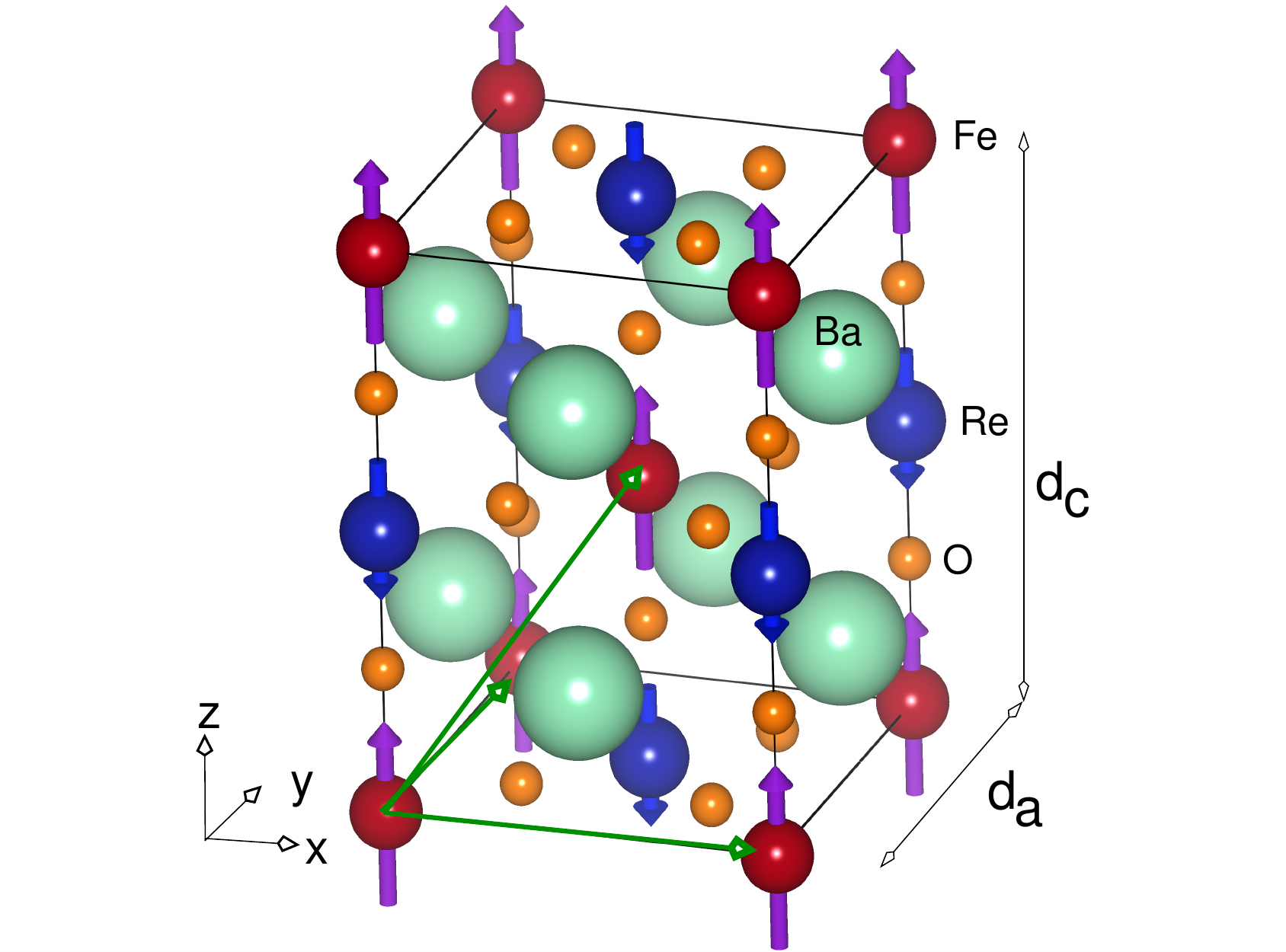}
    \caption{\label{Fig:xtal} Crystal structure of Ba$_2$FeReO$_6$ showing the choice of axes,
    the unit vectors for the elementary triclinic unit cell (green arrows), and magnetic moments in
    the ferrimagnetic ground state. Also shown is the enlarged
    body-centered tetragonal unit cell with lattice parameters $d_a$ and $d_c=d_a \sqrt{2}$.}
    \end{figure}

In this paper, we focus on metallic ordered DPs with mixed 3d/5d transition metal ions on the B/B' sites, 
specifically the Ba$_2$FeReO$_6$ material,\cite{Sleight1972,Prellier2000} with the structure as shown in Fig.\ref{Fig:xtal}.
we obtain the following main results. (i) We consider a 
model of the ordered double perovskite Ba$_2$FeReO$_6$
(see Fig.\ref{Fig:xtal}) retaining the relevant electronic states in the vicinity 
of the Fermi level. This model, after taking spin-orbit coupling as
well as correlations effects into account within a self-consistent mean field theory, is shown to reproduce previous
{\it ab initio} electronic structure results \cite{Jeon2010} in the ferrimagnetic ground state. 
Our model accounts for the dominant energy scales in this material: (a) the strong Hund's coupling on Fe,
the Hubbard repulsion on Re, and the Fe-Re charge transfer energy (all on the scale of $\sim 1$eV), (b) the strong spin-orbit coupling on 
Re ($\sim\!\!0.5$eV), and (c) the nearest neighbor Re-Fe 
hopping terms which leads to electron itinerancy ($\sim 0.3$eV). In addition, we include weaker terms such as inter-orbital mixing and 
second neighbor hopping which are required to reproduce the band degeneracies at high symmetry points 
in the Brillouin zone found in earlier {\it ab initio} studies. (ii) Our theory accounts semi-quantitatively
for the measured saturation magnetization \cite{Teresa2007}, as well as X-ray magnetic circular dichroism (XMCD) 
experiments which 
find a significant orbital contribution to the Re magnetization in
the ordered state.\cite{Azimonte2007,Winkler_NJP2009} (iii) Based on the orbital occupations in the magnetically ordered state, we predict a tetragonal
distortion, with c-axis compression accompanying magnetic order, in agreement with experimental data.\cite{Azimonte2007,Winkler_NJP2009} 
We also predict a specific doping
dependence to this orbital order and distortion which could be tested in future experiments. (iv) The strong correlations on Re, inferred 
from our study, lends partial support to earlier work
which showed that a local moment description of the ferrimagnetic state provides a reasonably good description of the 
magnetic dynamic structure factor obtained using inelastic neutron scattering experiments. \cite{Plumb_PRB2013} This
importance of strong correlation effects and local moment physics on
the 5d element is
in agreement with previous {\it ab initio} studies \cite{Das_PRB2011} that discussed the 
emergence of local moments of closely related Cr-based 3d/5d DPs Sr$_2$CrB'O$_6$
upon progressing through the series with B'$=$W,Re,Os.
(v) From our computed band dispersion, we show the appearance of
Weyl nodes in such metallic ferrimagnetic DPs. This is in line with the general understanding that in the presence of 
spin-orbit coupling, such Weyl nodes
are expected to be induced by breaking of time-reversal symmetry or
inversion symmetry.\cite{Burkov_PRB2011,Halasz_PRB2012} (vi) Using the Kubo formula for the spin-orbit coupled bands,
we find that Ba$_2$FeReO$_6$ itself appears to have only a small intrinsic anomalous Hall effect (AHE) in the 
ordered ferrimagnetic state at low temperature, but the AHE is significant in hole doped systems, and we speculate that it might
also be significant at intermediate temperatures below the ferrimagnetic $T_c$ in Ba$_2$FeReO$_6$.

Taking a broader viewpoint, Re-based layered quasi-two-dimensional oxides or heterostructures may be 
more strongly correlated than the three-dimensional DPs, and may lead to interesting Mott physics \cite{Chen2010,Chen2011}
beyond the iridates due to the local competition between interactions and spin-orbit coupling due to the d$^2$ 
configuration of Re$^{5+}$. Furthermore, one can carry out detailed inelastic neutron scattering studies in Re-based oxides, 
thus allowing for the possibility to explore the magnetism in more detail than in the iridates. 
This may prove to be useful in future studies of exotic variants of Re-based oxides.

\section{Model}
The simple charge counting for Ba$_2$FeReO$_6$ suggests Re$^{5+}$ and Fe$^{3+}$ valence states on the transition
metal ions. In this state, the five 3d-electrons on Fe are expected to be locked into a spin-$5/2$ moment due to strong
Hund's coupling in the half-filled d-shell. Here, we will treat this magnetic moment as a classical vector.
The two 5d-electrons in the Re $t_{2g}$ orbital are mobile, able to hop on and off the Fe sites subject to a charge transfer energy
$\Delta = {\cal E}_{\rm Fe}-{\cal E}_{\rm Re}>0$, and Pauli exclusion which constrains electrons arriving on Fe to be antiparallel to the 
direction of the local Fe moment. For a general direction of the Fe moment, $\vec F = (\sin\theta\cos\phi,\sin\theta\sin\phi,\cos\theta)$
at a given site,
we must project the added electrons onto the allowed direction to satisfy Pauli exclusion, locally 
setting $f_{\upa} = \sin\frac{\theta}{2} {\rm e}^{-i\phi/2} f$ and $f_\dna = -\cos\frac{\theta}{2} {\rm e}^{i\phi/2} f$, effectively 	``stripping''
the electron of its spin degree of freedom. Such models
have been proposed for other DP materials,\cite{Sarma_PRL2000,Sarma_PRB2001,Jackeli2003,Chattopadhyay_PRB2001,Phillips_PRB2003,Brey2006,Erten_SFMO_PRL2011,Das_PRB2011} 
and shown to capture the phenomenology 
of Sr$_2$FeMoO$_6$ including thermal phase transitions and disorder effects. \cite{Erten2011,ErtenPRB2013,MeeteiPRB2013}
However, most of these previous studies, with the notable exception of Ref.~\onlinecite{Das_PRB2011}
have ignored spin-orbit coupling effects, which are expected to be extremely important for 5d transition metal oxides. 

Our model
does not explicitly account for additional superexchange interactions between the Fe local moment and the emerging local
moments on the Re sites which is explicitly taken into account as a separate term in some previous studies (for example, Ref.~\onlinecite{Das_PRB2011});
however, we think such terms should emerge more naturally from an effective tight-binding model when strong correlations
are incorporated, as might be relevant to Mott insulating oxides like Sr$_2$CrOsO$_6$. Fe-Fe superexchange terms which
we omit, since they are not necessary to drive the ferrimagnetic state observed in Ba2FeReO6, may prove to be important in
understanding the complete magnetic phase diagram as a function of doping which is not addressed in this paper. However,
they are likely to be small given the Fe-Fe separation in the DP structure.
Further differences between the
results of Ref.~\onlinecite{Das_PRB2011} and our work stem from the fact that their model is for d$^3$ configuration of Cr, as 
opposed to our d$^5$ state on Fe; while both spin components of the itinerant electrons are permitted on Cr (since the
e$_g$ orbital is available), only one spin projection is allowed for itinerant electrons on Fe due to the Pauli exclusion.

\subsection{Non-interacting tight binding model}

The model describing Re electrons moving in the presence of Fe moments then takes the form
$H_0 =H_{\rm hop} + H_{\rm so} + H_{\rm ct}$. Here, the Hamiltonian $H_{\rm hop}$ describes intra-orbital 
hopping of electrons on the lattice, from Re to Fe (nearest-neighbor) and from Re to Re (next-neighbor), as
well as inter-orbital hopping of electrons between next-neighbor Re sites; $H_{\rm so}$ is the atomic spin-orbit
coupling on Re, projected to the $t_{2g}$ manifold, of strength $\lambda$; finally, $H_{\rm ct}$ describes the charge transfer energy
offset $\Delta$ between Re and Fe sites. For simplicity, we only focus on the case of a uniform magnetization on the Fe site,
assuming $(\theta,\phi)$
which describe the Fe moment to be site-independent;
it is straightforward to generalize our work to a nonuniform spatially varying magnetization. We use the simple
triclinic unit cell, with one Re and one Fe atom, as shown in Fig.\ref{Fig:xtal} to study the model Hamiltonian; however
in order to facilitate
a comparison with published {\it ab initio} electronic structure calculations, 
we will later assume a body-centered tetragonal unit cell containing two Re and two Fe atoms, 
with lattice constants $d_a\!\!=\!\!d_b\! \!=\!\! d_c/\sqrt{2}$ as shown in Fig.~\ref{Fig:xtal}, and use orthorhombic notation to 
plot the band dispersion of the eighteen bands in the Brillouin zone.
    
\begin{figure}[t]
    \includegraphics[scale=0.5]{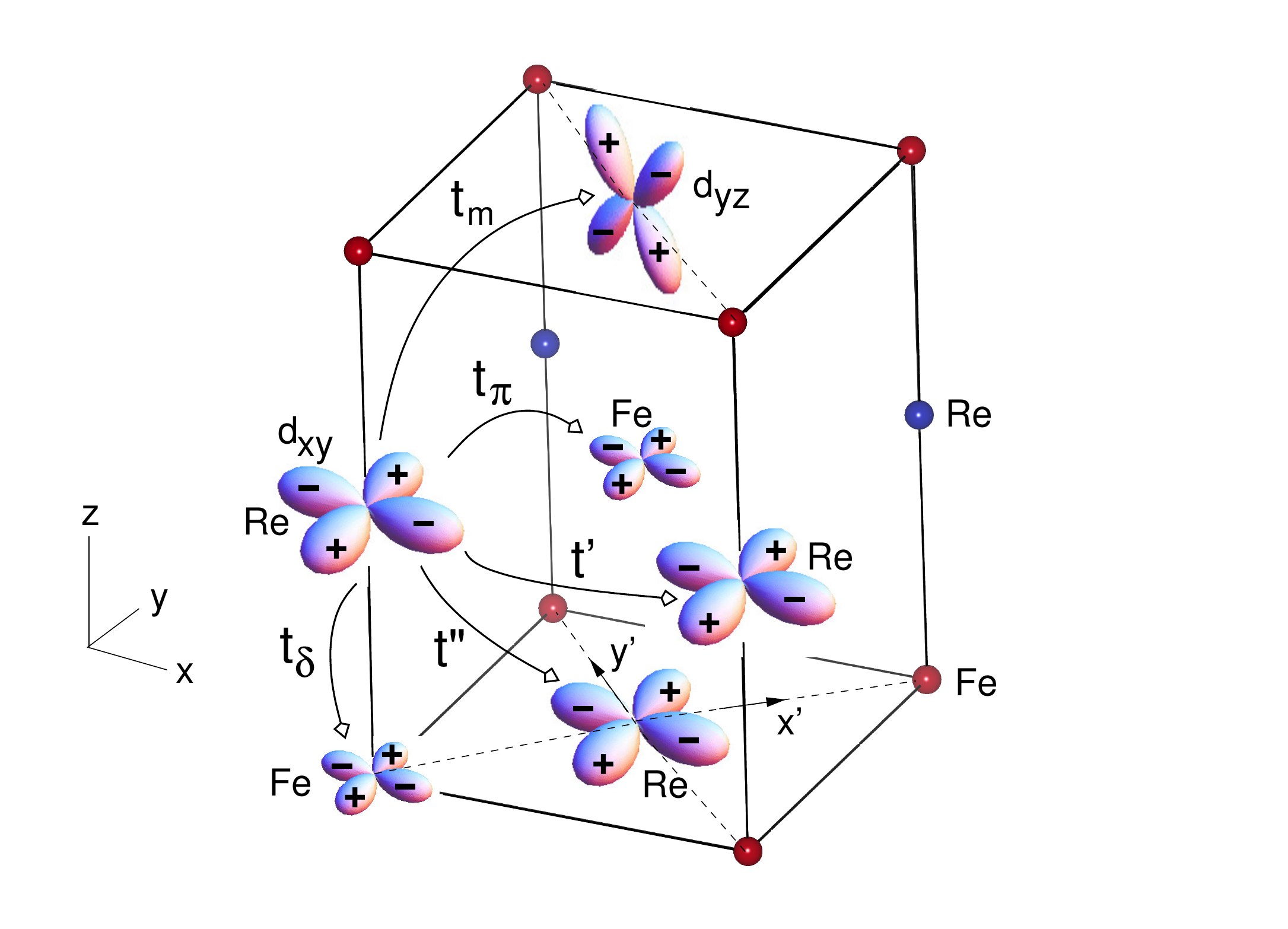}
    \caption{\label{Fig:hopping} Symmetry-allowed hopping matrix elements for double perovskites A$_2$BB'O$_6$ 
    (e.g., Ba$_2$FeReO$_6$), indicated for a few
    orbitals. $t_\pi,t_\delta$ are B-B' (Fe-Re) intraorbital hoppings, $t',t''$ are B'-B' (Re-Re) intraorbital hoppings, and $t_m$ denotes
    the interorbital B'-B' (Re-Re) hopping. All processes related to these by cubic symmetry are allowed. The Fe-Fe hoppings are 
    identical to Re-Re hoppings, but scaled by a factor $\alpha_f=0.5$. Also shown are the rotated axes (compass) for the tetragonal 
    unit cell of Ba$_2$FeReO$_6$, with $x'$-$y'$ (dashed lines) being the original cubic axes for defining the orbitals. }
    \end{figure}

We label the electrons on the Fe and Re sites by $f_{\ell}$ and 
$d_{\ell\sigma}$ respectively,
with $\ell \!\!=\!\! (1\! \equiv\! yz,2\! \equiv \! xz,3 \! \equiv \! xy)$ denoting the orbital, and $\sigma=\upa,\dna$ 
being the spin. The Hamiltonian takes the following form in momentum space, where we assume implicit
summation over repeated spin and orbital indices,
\bea
H_{\rm hop} \!\!\! &=&\!\!\!
\sum_{\bk} (\eta^\pdg_{\ell}(\bk) g^\pdg_\sigma(\theta,\phi) d^\dg_{\ell\sigma}(\bk) f^\pdg_{\ell}(\bk) + {\rm h.c.})  \nonumber\\
&+&\!\!\!\! \sum_{\bk} \! \epsilon^{\pdg}_{\ell}(\bk)(d^\dg_{\ell\sigma}(\bk) d^\pdg_{\ell\sigma}(\bk) 
+ \alpha_f f^\dg_{\ell}(\bk) f^\pdg_{\ell}(\bk)) \nonumber \\
&+& \!\!\!\!\!\!\! \sum_{\bk (\ell\neq\ell')} \!\!\!\!\!
\gamma^{\pdg}_{\ell\ell'}(\bk) (d^\dg_{\ell\sigma}(\bk) d^\pdg_{\ell' \sigma}(\bk)
\!+\! \alpha_f f^\dg_{\ell}(\bk) f^\pdg_{\ell'}(\bk)) \label{Hop} \\
H_{\rm so} 
\!\!\!&=&\!\!\! i \frac{\lambda}{2} \sum_{\bk} \varepsilon^\pdg_{\ell m n} \tau^n_{\sigma\sigma'} d^\dg_{\ell \sigma}(\bk) d^\pdg_{m \sigma'}(\bk) \\
H_{\rm ct} \!\!\!&=&\!\!\! \Delta \sum_{\bk} f^\dg_{\ell}(\bk) f^\pdg_{\ell}(\bk)
\eea
Here, in light of our previous discussion, we have only retained a single
spin projection on the Fe site, with
$g_\upa(\theta,\phi)= \sin\frac{\theta}{2} {\rm e}^{-i\phi/2}$ and $g_\dna(\theta,\phi) = -\cos\frac{\theta}{2} {\rm e}^{i\phi/2}$.
The various hopping processes are schematically illustrated in Fig.~\ref{Fig:hopping}.
The first term in $H_{\rm hop}$ describes nearest-neighbor intra-orbital hopping from Re to Fe, parameterized by
$t_\pi,t_\delta$.
The next two terms in $H_{\rm hop}$ characterize next-neighbor hopping processes, with the ratio of Fe-Fe
hoppings to Re-Re hoppings being $\alpha_f$; we will fix $\alpha_f =0.5$. While the
second term captures intra-orbital hopping between 
closest pairs of Re atoms or Fe atoms (parameterized by $t',t''$), the third
term captures inter-orbital hopping between closest pairs of Re atoms or Fe atoms
(parameterized by $t_m$). Many of these hopping processes ($t_\delta, t_m, t''$) have a small energy scale; however 
they are important to reproduce the band degeneracies found in {\it ab initio} calculations at high symmetry points in the
Brillouin zone.
The explicit momentum dependence of the dispersion coefficients appearing in $H_{\rm hop}$
is given in Appendix A.

\subsection{Interaction effects}
Electron-electron interactions are partially accounted for by $H_0$ in the previous section --- in part, by the charge
transfer gap $\Delta$, and, in part, by the implicit Hund's coupling which locks the Fe electrons into a high-spin state. 
However, electronic interactions on Re have been omitted in $H_0$. We
next include these local Hubbard interactions on Re.
The interaction Hamiltonian in the t$_{2g}$ orbitals of Re takes the form \cite{Fazekas}
\bea 
\!\!\!\! H_{\rm int} \!\!\!&=&\!\!\! U \sum_{i \ell \alpha} n^\pdg_{i \ell\upa}
n^\pdg_{i \ell\dna} + (U \!-\! 5 \frac{J_H}{2}) \sum_{\ell<\ell'} n^\pdg_{i \ell}
n^\pdg_{i \ell'} \nonumber \\
\!\!\! &-&\!\!\! 2 J_H \sum_{\ell< \ell'} \vec S^\pdg_{i \ell}
\cdot \vec S^\pdg_{i \ell'} + J_H \sum_{\ell \neq \ell'} d^\dg_{i \ell\upa}
d^\dg_{i \ell\dna} d^\pdg_{i \ell' \dna} d^\pdg_{i \ell' \upa}
\eea
where $i$ labels the Re sites, and $\vec S_{i\ell} = \frac{1}{2} d^\dg_{i\ell\alpha} \vec \sigma_{\alpha\beta} d^\pdg_{i\ell\beta}$ is the spin
at site $i$ in orbital $\ell$. We wish to then study the full Hamiltonian $H = H_0 + H_{\rm int}$. 
For simplicity, we only retain only the dominant intra-orbital Coulomb repulsion, treating it at mean field (Hartree) level, as
\bea
\!\!\!\!\! H_{\rm int} \!\!\!&\approx&\!\!\! U \! \sum_{i\ell}\!\! \left[\! \frac{\rho_\ell}{2} (n_{i\ell \upa} \!+\! n_{i\ell\dna}) \!-\! 
2 \vec{m}_{\ell}\cdot \vec S_{i\ell} \!-\! \frac{\rho^2_\ell}{4} \!+\! \vec m_{\ell} \! \cdot \! \vec m_{\ell} \! \right]
\eea
where $\rho_\ell = \la n_{i\ell\upa} + n_{i\ell\dna}\ra$, $\vec m_{\ell} = \la \vec S_{i\ell} \ra$, and we set $\vec m_\ell = - m_\ell (\sin\theta\cos\phi,\sin\theta\sin\phi,\cos\theta)$,
with $m_\ell > 0$, so that $\vec m_\ell$ is anti-parallel to the Fe moment $\vec F$. 
For simplicity, we only focus on the case $\theta=\phi=0$, so the Fe sites can
only accommodate itinerant spin-$\dna$ electrons. We
then numerically determine $m_\ell$ and $\rho_\ell$ in a self-consistent fashion, using the non-interacting ground state
as the starting point for the iterative solution, while ensuring that the choice of the chemical potential lead to 
a total of two electrons per unit cell (i.e., per Re atom).
Such a mean field treatment of electron-electron interactions
does not capture all aspects of the strong correlation physics, e.g. bandwidth renormalization and mass
enhancement. Nevertheless, recognizing this caveat, we use
the self-consistent solution of the mean field equations to study the effects of interactions
and spin-orbit coupling on the reorganization of the nine electronic bands, compare the
physical properties with experimental results, and make qualitative predictions for future experiments.

\section{Physical properties}

We begin by discussing the effect of electronic correlations in the DPs in the absence of spin-orbit coupling. We show that such correlation effects 
appear to be crucial to stabilize a half-metallic state with complete polarization in the 5d perovskites, due to the large second-neighbor Re-Re 
hopping which otherwise prevents a half-metallic state. We then turn to the effect of spin-orbit coupling, and show that it reorganizes the band
structure, yielding results which are in reasonable agreement with previous {\it ab initio} electronic structure studies.\cite{Jeon2010} 
(As pointed out earlier, the
band dispersions discussed below are plotted using the orthorhombic notation with an enlarged unit cell containing two Fe and two Re atoms,
leading to eighteen electronic bands instead of nine.) Finally, we compare the mean field result for the saturation magnetization with experiments, 
and the spin and orbital magnetization on the Re site with previous XMCD data, and discuss other physical
properties such as tetragonal lattice distortion and predictions for the AHE. Throughout this discussion, we will assume a ferromagnetic order of
the Fe moments - a more complete study of the magnetic phase diagram as a function of doping and temperature will be the subject of future
numerical investigations.

\begin{figure}[tb]
\includegraphics[scale=0.53]{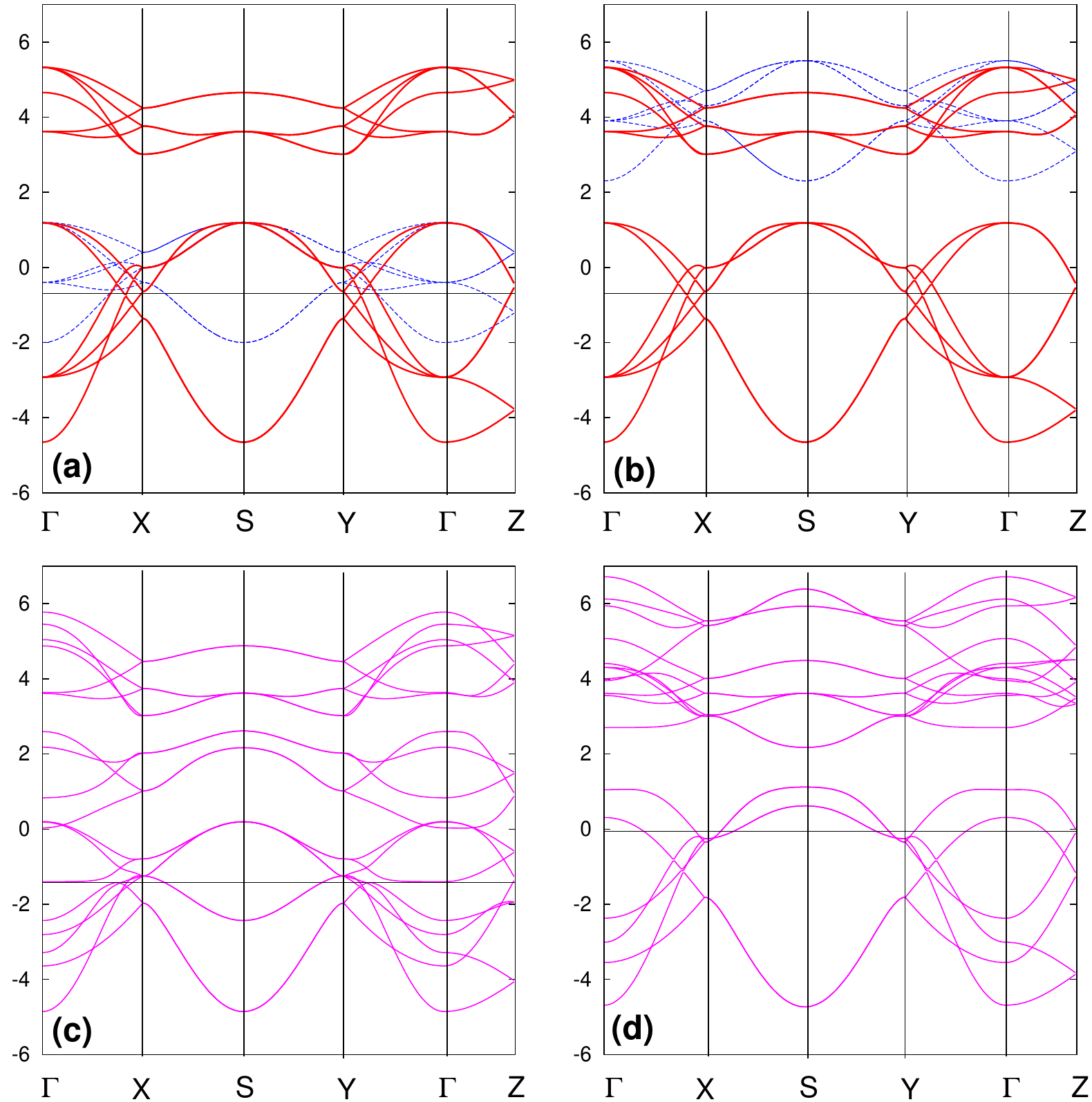}
    \caption{\label{Fig:band} Band dispersion in the orthorhombic notation for the Re and Fe electronic states for different choices of Hubbard interaction $U$ 
    and spin-orbit coupling $\lambda$, with energy on the $y$-axis in units of $t_\pi$. The solid black line indicates the chemical potential. 
    For no spin-orbit coupling, (a) $U=0$ and $\lambda=0$ and (b) $U=8 t_\pi$ and $\lambda=0$, we find
    decoupled spin-$\dna$ (red, solid) and spin-$\upa$ (blue, dashed) states. Comparing (a) and (b), we see that
    correlations on Re push the spin-$\upa$ states to higher energy, leading to the stabilization of a half-metal 
    ground state. A nonzero spin-orbit coupling, 
    (c) $U=0$ and $\lambda = 2 t_\pi$, and (d) $U=8 t_\pi$ and $\lambda=2 t_\pi$, leads to
    mixed-spin states and splits degeneracies, but for a physically reasonable value $U=8 t_\pi$ 
    preserves significant spin polarization $\sim 90\%$ for states at the Fermi level.}
\end{figure}

\subsection{Correlations stabilize a half-metal}
If we ignore Re correlations entirely, setting $U=0$, and also ignore spin-orbit coupling by setting $\lambda=0$, the band 
structure shown in Fig.\ref{Fig:band}(a) has decoupled spin-$\upa$ and spin-$\dna$ bands. The twelve spin-$\dna$
bands corresponding to electrons which can delocalize on Re and Fe. By contrast, the six spin-$\upa$ bands corresponds to purely 
Re states. Working in units where $t_\pi=1$, we find that to make a reasonable comparison with the {\it ab initio} calculations,
we have to choose a significant $t'=0.3$ (Re-Re hopping), but all other hoppings can be assumed to be small; for simplicity,
we fix $t_\delta=t''=t_m=0.1$. Finally, we have to assume a moderate charge transfer energy $\Delta=3$ which splits the spin-$\dna$
states into two groups: $6$ lower energy Re-Fe hybridized spin-$\dna$ states (dominant Re character) which form a {\it broad} band, and $6$ 
higher energy dominantly Re-Fe hybridized spin-$\dna$ states (dominant Fe character) which form a {\it narrow}
band. Finally, the remaining $6$ Re-$\upa$ states form a {\it narrow} dispersing band, crossing the chemical potential and overlapping
in energy with the broad spin-$\dna$ band.
For $U=0$, the system thus contains both spin states at the Fermi level. When we incorporate a Hubbard repulsion $U=8 t_\pi$
at mean field level, we see from Fig.~\ref{Fig:band}(b) that its main effect is to self-consistently shift the spin-$\upa$ bands higher in energy,
leaving only spin-$\dna$ states at the Fermi level. The resulting band dispersion is in reasonably good agreement with LDA+U 
calculations. Although we have not attempted a detailed quantitative fitting to the LDA+U band structure, the features  noted below
are robust.
(i) A rough comparison 
with the overall bandwidth in the {\it ab initio} calculations without spin-orbit coupling \cite{Jeon2010} suggests that $t_\pi \approx 330$meV.
This is somewhat larger than
estimates for Sr$_2$FeMoO$_6$ in the literature \cite{Sarma_PRL2000,Sarma_PRB2001,Erten_SFMO_PRL2011} ($\sim \!\! 270$ meV). 
(ii) We estimate the interaction energy scale on Re to be $U \approx 2.5$eV, smaller by a factor of 
two compared with typical values for 3d transition metals. (iii) There is a significant Re-Re hopping, $t'/t_\pi \sim 0.3$, 
we need to include in order to be able to capture the bandwidths of the spin-$\upa$ and spin-$\dna$ bands. All these 
observations are reasonable given the more extended nature of Re 
orbitals when compared with 3d or 4d transition metal ions. The presence of appreciable Re-Re hoppings has been pointed
out in previous work, \cite{Gopalakrishnan_PRB2000,Chattopadhyay_PRB2001} although they did not take correlation effects on
Re into account. More recent work has also arrived at similar conclusions regarding significant Re-Re hoppings.\cite{Das_PRB2011}

{\it To summarize, we have obtained a tight-binding description including interactions of DPs with spin-orbit coupling. In 
contrast to 3d/4d DP materials like Sr$_2$FeMoO$_6$, we find that 3d/5d DPs have a significant second neighbor
hopping; strong correlations 
on the 5d element (Re) therefore play a crucial role in stabilizing a half-metallic ground state in the 3d/5d DPs}.

\subsection{Spin-orbit coupling: Band reconstruction, spin/orbital magnetization, and comparison with
magnetization and XMCD experiments}
We next turn to the effect of incorporating both spin-orbit coupling and Hubbard interactions on Re, solving the
mean field equations in case of a nonzero $U$.
From Fig.\ref{Fig:band}(c) and (d), where we have set $\lambda=2 t_\pi$ ($\sim\!\! 660$meV for our estimated $t_\pi$),
we see that spin-orbit coupling
clearly eliminates the degeneracies occurring at the $\Gamma$-point for $\lambda=0$. It also
significantly reconstructs the dispersion of the eighteen bands, leading to reasonably good agreement with
published {\it ab initio} calculations which include spin-orbit coupling.\cite{Jeon2010} In the next section, we will discuss the resulting appearance
of Weyl nodes in the band dispersion and the intrinsic anomalous Hall effect in the ordered state. Here, we will use
the mean field solution to estimate the average Fe valence, the Fe ordered moment, 
and the spin and orbital contributions to the Re moment. In the ground state with correlations, we find that the average valence of Fe shifts from the
naive charge counting value Fe$^{3+}$ to Fe$^{2.6+}$, and the Fe moment is lowered to an effective value ${\cal F}_z \approx 2.3$ (corresponding
to $4.6 \mu_B$). Quantum
spin fluctuations beyond the mean field result might further slightly suppress this value. On Re, we find an ordered spin moment 
${\cal S}_z  \approx 0.78$ and an orbital moment ${\cal L}_z \approx 0.48$; taking the g-factor into account, and undoing the sign change of the orbital 
angular momentum which appears upon 
projection to the $t_{2g}$ Hamiltonian, this implies a ratio of magnetic moments $\mu^{\rm orb}_{\rm Re}/\mu^{\rm spin}_{\rm Re} \approx -0.31$, remarkably 
close to the experimentally measured XMCD result $\approx -0.29$. We find that the actual value of the spin 
magnetic moment, $\mu^{\rm spin}_{\rm Re} \approx 1.56 \mu_B$,  is larger than the experimentally reported XMCD value
$\approx 1.08 \mu_B$. This
discrepancy might be partly due to the fact that (i) the 
experimental results are on powder samples, and hence might appear to be smaller simply due to averaging over grain orientations, and (ii) the method
to extract the individual spin or orbital magnetic moments relies on additional assumptions, while the ratio is apparently more 
reliable.\cite{Azimonte2007} We must contrast these results with the case where we ignore Re correlations entirely; in that case, the Fe moment is not much affected,
${\cal F}_z^{U=0} \approx 2.4$,
but the Re moments are strongly suppressed, yielding ${\cal S}_z^{U=0} \approx 0.15$ and an orbital moment ${\cal L}_z^{U=0} \approx 0.09$ which
would lead to a much smaller $\mu^{\rm spin}_{\rm Re}(U=0) \approx 0.3 \mu_B$ than is experimentally estimated, as well as a much larger 
saturation magnetization, $4.6 \mu_B$, 
than the measured value \cite{Prellier2000,Teresa2007} which is $\approx 3.2$-$3.3 \mu_B$. Our estimates 
in the presence of correlations, by contrast, yield $m_{\rm sat} \approx 3.5 \mu_B$, in much better agreement with the data. Finally, we
use our solution to estimate the polarization, defined as the
degree of magnetization for states near the Fermi level. We find that while the correlated half-metal state in the absence 
of spin-orbit coupling exhibits (obviously) $100\%$ polarization, using $\lambda=2t_\pi$ reduces the polarization to $\sim\! 90\%$. However,
if we only take spin-orbit coupling into account and ignore strong correlations, the states near the Fermi level are nearly unpolarized.

{\it In 3d/5d DP materials, spin orbit coupling and strong correlations are both crucial to obtain the experimentally observed spin and 
orbital magnetization and their locking, and to explain the experimentally observed saturation magnetization and XMCD signal. 
Spin-orbit coupling leads to a slight decrease of the correlation-induced spin polarization at the Fermi level.}

\subsection{Orbital order, tetragonal distortion in ferrimagnetic state, and doping dependence}
In the converged mean field state, with the magnetization along the $z$-axis, we find that the density on Re in the three orbitals are 
different, with $\rho_{xy} \! \approx \! 0.60$ and $\rho_{xz} \! = \! \rho_{yz} \! \approx \! 0.53$. This orbital imbalance is induced in the
$z$-ferrimagnetic state due the
spin-orbit coupling. The larger extent of the $xy$-orbital 
in the $xy$-plane, compared with its smaller extent along the $z$-direction, implies that this orbital charge imbalance would lead to a
tetragonal distortion of the lattice, to occur
coincident with ferrimagnetic ordering and with a shrinking of the c-axis, as has indeed been observed to occur experimentally.
The precise extent of this distortion, which is observed \cite{Azimonte2007} to be $\sim 0.1\%$, depends on details such as the
lattice stiffness, and is beyond the scope of our calculation. 

\begin{figure}[t]
  \includegraphics[scale=0.2]{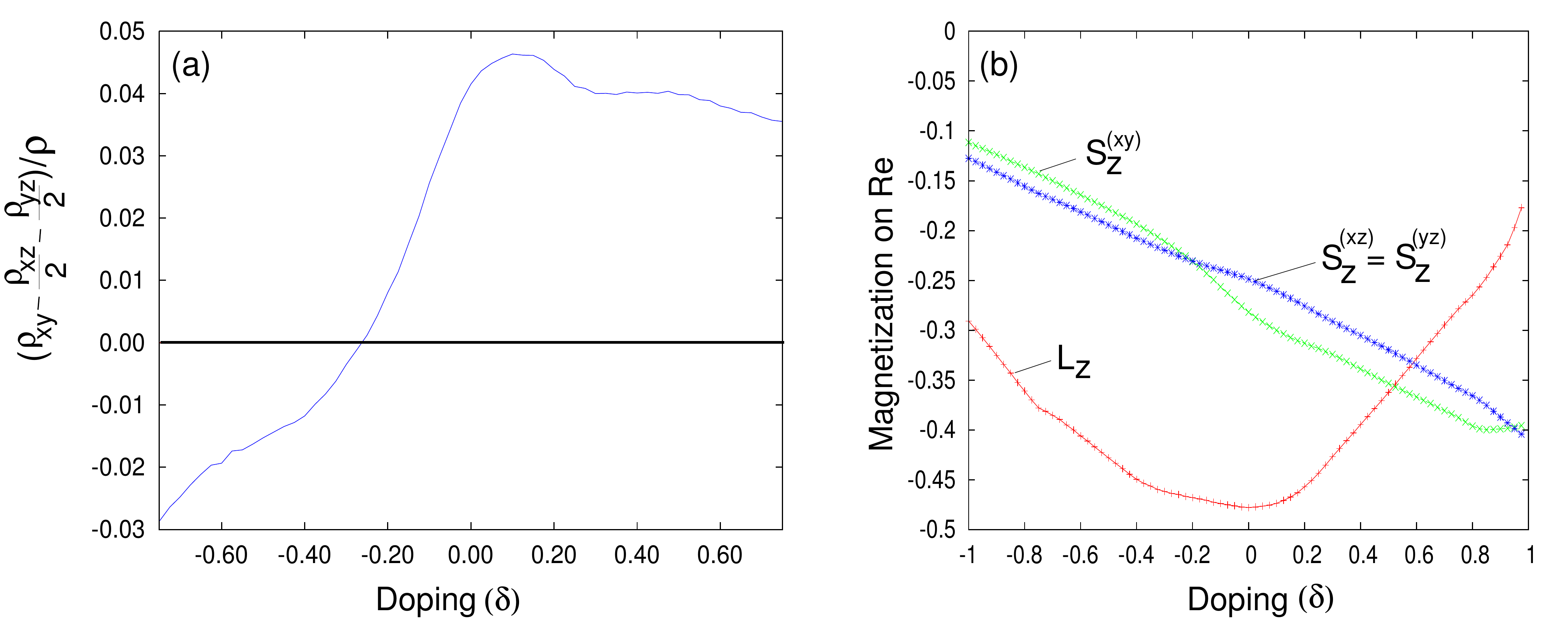}
  \caption{\label{Fig:orborder} (a) Relative orbital occupancy in the ferrimagnetic state of Ba$_2$FeReO$_6$ incorporating
    mean-field interactions and strong spin-orbit coupling at a doping of $\delta$ excess electrons per Re. The parameters 
    used are the same as those for Fig.~\ref{Fig:band}(b), namely $U=8t_\pi$ and $\lambda=2t_\pi$. This orbital order implies 
    a tetragonal distortion of the lattice, with c-axis compression for $\delta \gtrsim -0.25$, and c-axis elongation for $\delta \lesssim -0.25$.
    (b) Doping dependence of orbital magnetization on Re, and the various orbital components of the spin magnetization on Re.}
\end{figure}

When we solve the self-consistent
equations at various dopings $\delta$ (excess electrons per Re) assuming persistent ferrimagnetic order,
the extent of this orbital imbalance, characterized by a tetragonal order parameter
$\eta_{\rm tet} = \frac{1}{\rho}(\rho_{xy}-\rho_{xz}/2-\rho_{yz}/2)$, changes systematically 
as shown in Fig.~\ref{Fig:orborder}(a).
Light electron doping leads to a slightly larger orbital population imbalance and
should enhance the $c$-axis compression, while a larger electron doping leads to a gradual decrease 
of $\eta_{\rm tet}$. Hole doping beyond
$\gtrsim \! 0.25$ holes/Re leads to $\eta_{\rm tet} < 0$, which should cause elongation along the $c$-axis. The
spin contribution to the magnetization on Re, arising from the different orbitals, also shows a similar doping trend as seen from Fig.~\ref{Fig:orborder}(b),
while the orbital contribution to the magnetization on Re has the largest magnitude at zero doping.
These results could be possibly be explored experimentally
by partially substituting Ba by trivalent La (electron doping), or by Cs or other monovalent ions (hole doping).

{\it Thus, in 3d/5d DP materials, spin orbit coupling and the ferrimagnetic order of itinerant electrons leads to orbital
ordering. This, in turn, should lead to a compression along the $c$-axis, consistent with the experimentally observed tetragonal
distortion, and we predict a specific doping dependence to this structural distortion.}

\subsection{Doping-dependent anomalous Hall effect}
\begin{figure}[t]
  \includegraphics[scale=0.42]{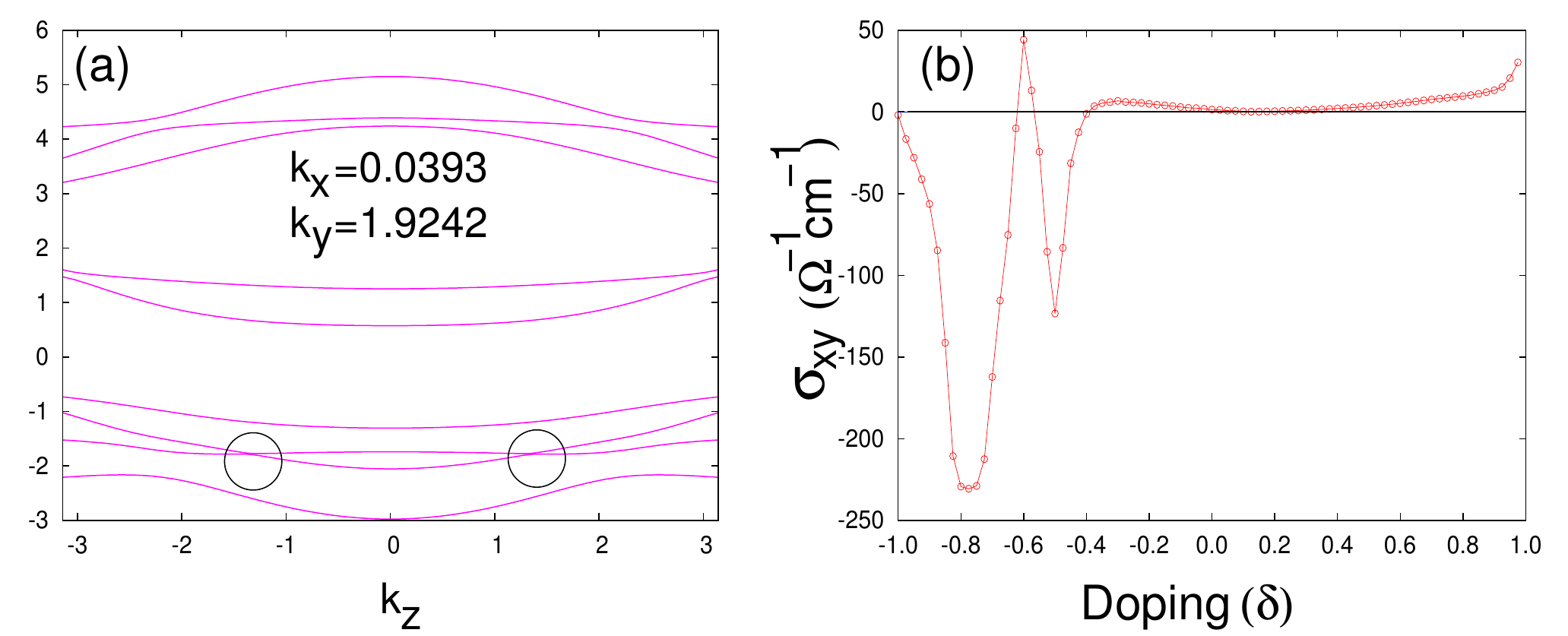}
  \caption{\label{Fig:weyl} (a) Band dispersion of Ba$_2$FeReO$_6$ in the presence of 
    interactions and spin-orbit coupling plotted along $k_z$ at fixed $(k_x\!=\!0.0393,k_y\!=\!1.9242)$,
    showing the nine bands and a pair of Weyl nodes (with energy on $y$-axis in units of $t_\pi$). The parameters used are the same as those for Fig.~\ref{Fig:band}(b),
    namely $U\!=\! 8t_\pi$ and $\lambda\! =\! 2t_\pi$. (b) Doping dependence of the intrinsic anomalous Hall conductivity $\sigma_{xy}$.}
\end{figure}

We next turn to the intrinsic AHE in the ferrimagnetic state of such 3d/5d DPs. As pointed out in recent work,
for pyrochlore iridates with all-in-all-out order under uniaxial pressure,\cite{YRan_PRB2011} as well the ferromagnetic infinite-layer
ruthenate SrRuO$_3$, \cite{Burkov_WeylAHE} this
intrinsic AHE contains two contributions: (i) a surface contribution arising 
from Fermi arc states  \cite{Wan_PRB2011} associated with Weyl nodes in the dispersion, and (ii) a bulk contribution from
carriers near the Fermi surface.
A pair of such Weyl nodes for Ba$_2$FeReO$_6$ is shown in Fig.~\ref{Fig:weyl} obtained from the interacting band dispersion.
\footnote{The full set of Weyl
nodes - their location, charges, and dependence on the direction of the magnetization vector - will be discussed elsewhere.
(A. M. Cook, A. A. Burkov, and A. Paramekanti, work in progress)}

Both contributions to the intrinsic AHE are captured by the momentum-dependent Berry curvature \cite{Niu_PRB1999,RMP_AHE} 
of the spin-orbit coupled bands, which is, in turn, obtained from the Kubo formula
\bea
\sigma_{xy} = e^2 \hbar \int \frac{d^3\bk}{(2\pi)^3} \sum_{m \neq n} \frac{f(\varepsilon_{\bk m})-f(\varepsilon_{\bk n})}
{(\varepsilon_{\bk m}-\varepsilon_{\bk n})^2}{\rm Im}(v^x_{m n} v^y_{n m}).
\eea
Here, $\varepsilon_{\bk m}$ is the single-particle energy at momentum $\bk$ and band $m$, 
$v^\alpha = \frac{1}{\hbar}(\partial H_{\rm mf}/\partial k_\alpha)$ are components of the velocity operator 
with $H_{\rm mf}$ being the self-consistently determined mean-field Hamiltonian matrix, and
$f(.)$ is the Fermi function. \footnote{An equivalent route to computing $\sigma_{xy}$ is to view the 3D band
dispersion $\varepsilon_{\bk\ell}$ as a sequence of 2D band structures parameterized by the momentum $k_3$ along one 
direction,\cite{Burkov_PRL2011,Burkov_WeylAHE} writing it as $\varepsilon_\ell^{[k_3]} (k_1,k_2)$. Each such 2D band
can have a momentum dependent Berry curvature and a
nonzero Chern number, thus yielding a quantum Hall insulator for a filled band. 
Weyl nodes, which act as `monopoles' in momentum
space - sources or sinks of integer quanta of Berry flux \cite{Wan_PRB2011,YRan_PRB2011,Burkov_PRL2011} -
correspond to quantum Hall transitions in momentum space where the Chern number 
jumps as a function of $k_3$.
Integrating the Berry curvature, obtained by using gauge invariant plaquette products of wavefunction overlaps,\cite{Hatsugai_JPSJ2005,Burkov_WeylAHE} 
over all the $k_3$ slices yields the total $\sigma_{xy}$.}
From the mean field solution corresponding to two electrons per Re, as appropriate for Ba$_2$FeReO$_6$, we find that $\sigma_{xy}$ at
zero temperature
is small, $\sigma_{xy} \! \sim\!  10^{-3} \frac{e^2}{\hbar d_c}$ where $d_c \approx 8 \AA$ is the lattice constant in Fig.~\ref{Fig:xtal}. This translates into
$\sigma_{xy} \sim 5 \Omega^{-1}$cm$^{-1}$.

In order to explore $\sigma_{xy}$ over a larger space of parameters, we consider its variation with doping.
Rather than simply shifting the  chemical potential, we solve the Hartree mean field equations 
over a range of electron densities, and then compute $\sigma_{xy}$ in the resulting self-consistent band structure.
We find that electron doping does not significantly enhance the AHE, but a hole doping of about $0.5$-$0.8$ holes/Re leads to a 
larger AHE $\sigma_{xy}\! \sim \! - 100  \Omega^{-1}$cm$^{-1}$ to $-250 \Omega^{-1}$cm$^{-1}$. 
Even this significant AHE is small  in natural units ($\sim 0.1\frac{e^2}{\hbar d_c}$) at $T=0$, which we attribute to the large spin polarization 
in the completely ordered ferromagnet. It is possible that the AHE is a non-monotonic function of temperature, peaked at some 
intermediate temperature below the magnetic $T_c$ even in the undoped compound.

{\it Thus, in 3d/5d DP materials, spin orbit coupling and the ferrimagnetic order is expected to lead to an intrinsic 
AHE. The AHE appears likely to be larger for hole doped systems compared to an expected small value for Ba$_2$FeReO$_6$ and is
likely, in  Ba$_2$FeReO$_6$, to be peaked at intermediate temperatures below $T_c$.}

\section{Conclusion}
We have obtained a tight-binding description of the metallic DPs, including spin-orbit coupling and strong correlation effects. 
Although we have here only applied it to Ba$_2$FeReO$_6$, 
finding good agreement with a broad variety of experiments and with electronic structure calculations,
our work should
be broadly applicable to other 3d/4d and 3d/5d DP materials as well. Our finding that strong correlation effects
are needed to explain many of the experimental observations also lends partial justification to our previous theoretical
work which modelled the measured spin wave spectrum using a local moment model. 
Further theoretical work is needed to study the thermal fluctuation effects of the Fe moments, clarify what factors control the 
doping dependence of $\sigma_{xy}$, and to separate the bulk and surface contributions to the AHE. Furthermore, it would 
be useful to investigate if ferrimagnetic order in fact survives over a wide range of doping using an unbiased numerical 
approach. In future experiments, it would be useful to test our 
predictions for the doping dependence of the structural distortion and the AHE. 
Given that most DP materials are in the form of powder samples, measuring
the AHE and separating the intrinsic contribution from extrinsic contributions would be experimentally challenging; nevertheless
systematic doping studies of the various properties of such 3d/5d DPs would be valuable.
Finally, it appears 
extremely important to find ways to synthesize bulk single crystals or  high quality thin films of such DP materials
which would greatly open up the exploration of their 
physical properties and applications.

\bigskip

\acknowledgments
This research was supported by NSERC of Canada.
We acknowledge useful discussions with Anton Burkov, Patrick Clancy, Young-June Kim, Priya Mahadevan, Kemp Plumb,
Mohit Randeria, Nandini Trivedi, and Roser Valenti. AP acknowledges the support and hospitality of the Max-Planck-Institut
f\"ur Physik komplexer Systeme (Dresden), and discussions with the participants of the SPORE13 workshop,
where part of this work was completed.

\appendix

\section{Tight binding coefficients}
When we work with the triclinic unit cell, there is one Fe atom and one Re
atom in each unit cell. Going to momentum space, the coefficients of the tight-binding hopping Hamiltonian 
$H_{\rm hop}$ in Eq.~\ref{Hop} have intra-orbital terms given by
\bea
\!\! \epsilon_{xy} \!\!&=&\!\! -2 t' (\cos k_x a \!+\! \cos k_y a) \\
\!\!&+&\!\! 8 t'' \cos (\frac{k_x a}{2}) \cos (\frac{k_y a}{2}) \cos (\frac{k_z c}{2}) \\
\!\! \epsilon_{xz} \!\!&=&\!\! 2 t'' (\cos k_x a\! +\! \cos k_y a) \!+\! 4 t'' \!\! \cos(\frac{k_x a \!-\! k_y a}{2}) \cos(\frac{k_z c}{2}) \nonumber \\
\!\!&-&\!\! 4 t' \cos(\frac{k_x a \!+\! k_y a}{2}) \cos \frac{k_z c}{2}\\
\!\! \epsilon_{yz} \!\!&=&\!\! 2 t'' (\cos k_x a \!+\! \cos k_y a) \!+\! 4 t'' \!\! \cos(\frac{k_x a \!+\! k_y a}{2}) \cos(\frac{k_z c}{2}) \nonumber \\
\!\!&-&\!\! 4 t' \cos(\frac{k_x a\!-\! k_y a}{2}) \cos \frac{k_z c}{2},
\eea
and
\bea
\!\!\!\!\eta_{xy}  &=& 4 t_\pi  \cos \frac{k_x a}{2} \cos \frac{k_y a}{2} - 2 t_\delta \cos\frac{k_z c}{2} \\ 
\!\!\!\!\eta_{xz} &=& 2 t_\pi\! \cos (\frac{k_x a\!+\! k_y a}{2}) + 2 t_\pi \cos \frac{k_z c}{2} \\
&-& 2 t_\delta\! \cos (\frac{k_x a \! -\! k_y a}{2})  \\
\!\!\!\!\eta_{yz} &=& 2 t_\pi\! \cos (\frac{k_x a\!-\! k_y a}{2}) + 2 t_\pi \cos \frac{k_z c}{2} \\
&-& 2 t_\delta\! \cos (\frac{k_x a \! + \! k_y a}{2}).
\eea
The intra-orbital terms take the form
\bea
\gamma_{xz,yz} &=& -2 t_m (\cos k_x a - \cos k_y a) \\
\gamma_{xy,yz}&=& - 4 t_m \sin(\frac{k_x a + k_y a}{2}) \sin\frac{k_z c}{2}\\
\gamma_{xy,xz} &=& 4 t_m \sin(\frac{k_x a - k_y a}{2}) \sin\frac{k_z c}{2}.
\eea


\begin{thebibliography}{44}
\expandafter\ifx\csname natexlab\endcsname\relax\def\natexlab#1{#1}\fi
\expandafter\ifx\csname bibnamefont\endcsname\relax
  \def\bibnamefont#1{#1}\fi
\expandafter\ifx\csname bibfnamefont\endcsname\relax
  \def\bibfnamefont#1{#1}\fi
\expandafter\ifx\csname citenamefont\endcsname\relax
  \def\citenamefont#1{#1}\fi
\expandafter\ifx\csname url\endcsname\relax
  \def\url#1{\texttt{#1}}\fi
\expandafter\ifx\csname urlprefix\endcsname\relax\def\urlprefix{URL }\fi
\providecommand{\bibinfo}[2]{#2}
\providecommand{\eprint}[2][]{\url{#2}}

\bibitem[{\citenamefont{Serrate et~al.}(2007)\citenamefont{Serrate, Teresa, and
  Ibarra}}]{Serrate_Review_JPCM2007}
\bibinfo{author}{\bibfnamefont{D.}~\bibnamefont{Serrate}},
  \bibinfo{author}{\bibfnamefont{J.~M.~D.} \bibnamefont{Teresa}},
  \bibnamefont{and} \bibinfo{author}{\bibfnamefont{M.~R.}
  \bibnamefont{Ibarra}}, \bibinfo{journal}{Journal of Physics: Condensed
  Matter} \textbf{\bibinfo{volume}{19}}, \bibinfo{pages}{023201}
  (\bibinfo{year}{2007}).

\bibitem[{\citenamefont{Kobayashi et~al.}(1998)\citenamefont{Kobayashi, Kimura,
  Sawada, Terakura, and Tokura}}]{Kobayashi1998}
\bibinfo{author}{\bibfnamefont{K.~I.} \bibnamefont{Kobayashi}},
  \bibinfo{author}{\bibfnamefont{T.}~\bibnamefont{Kimura}},
  \bibinfo{author}{\bibfnamefont{H.}~\bibnamefont{Sawada}},
  \bibinfo{author}{\bibfnamefont{K.}~\bibnamefont{Terakura}}, \bibnamefont{and}
  \bibinfo{author}{\bibfnamefont{Y.}~\bibnamefont{Tokura}},
  \bibinfo{journal}{Nature} \textbf{\bibinfo{volume}{395}},
  \bibinfo{pages}{677} (\bibinfo{year}{1998}).

\bibitem[{\citenamefont{Sarma et~al.}(2000)\citenamefont{Sarma, Mahadevan,
  Saha-Dasgupta, Ray, and Kumar}}]{Sarma_PRL2000}
\bibinfo{author}{\bibfnamefont{D.~D.} \bibnamefont{Sarma}},
  \bibinfo{author}{\bibfnamefont{P.}~\bibnamefont{Mahadevan}},
  \bibinfo{author}{\bibfnamefont{T.}~\bibnamefont{Saha-Dasgupta}},
  \bibinfo{author}{\bibfnamefont{S.}~\bibnamefont{Ray}}, \bibnamefont{and}
 \bibinfo{author}{\bibfnamefont{A.}~\bibnamefont{Kumar}},
  \bibinfo{journal}{Phys. Rev. Lett.} \textbf{\bibinfo{volume}{85}},
  \bibinfo{pages}{2549} (\bibinfo{year}{2000}).

\bibitem[{\citenamefont{Saha-Dasgupta and Sarma}(2001)}]{Sarma_PRB2001}
\bibinfo{author}{\bibfnamefont{T.}~\bibnamefont{Saha-Dasgupta}}
  \bibnamefont{and} \bibinfo{author}{\bibfnamefont{D.~D.} \bibnamefont{Sarma}},
  \bibinfo{journal}{Phys. Rev. B} \textbf{\bibinfo{volume}{64}},
  \bibinfo{pages}{064408} (\bibinfo{year}{2001}).

\bibitem[{\citenamefont{Jackeli}(2003)}]{Jackeli2003}
\bibinfo{author}{\bibfnamefont{G.}~\bibnamefont{Jackeli}},
  \bibinfo{journal}{Phys. Rev. B} \textbf{\bibinfo{volume}{68}},
  \bibinfo{pages}{092401} (\bibinfo{year}{2003}).

\bibitem[{\citenamefont{Chattopadhyay and
  Millis}(2001)}]{Chattopadhyay_PRB2001}
\bibinfo{author}{\bibfnamefont{A.}~\bibnamefont{Chattopadhyay}}
  \bibnamefont{and} \bibinfo{author}{\bibfnamefont{A.~J.}
  \bibnamefont{Millis}}, \bibinfo{journal}{Phys. Rev. B}
  \textbf{\bibinfo{volume}{64}}, \bibinfo{pages}{024424}
  (\bibinfo{year}{2001}).

\bibitem[{\citenamefont{Phillips et~al.}(2003)\citenamefont{Phillips,
  Chattopadhyay, and Millis}}]{Phillips_PRB2003}
\bibinfo{author}{\bibfnamefont{K.}~\bibnamefont{Phillips}},
  \bibinfo{author}{\bibfnamefont{A.}~\bibnamefont{Chattopadhyay}},
  \bibnamefont{and} \bibinfo{author}{\bibfnamefont{A.~J.}
  \bibnamefont{Millis}}, \bibinfo{journal}{Phys. Rev. B}
  \textbf{\bibinfo{volume}{67}}, \bibinfo{pages}{125119}
  (\bibinfo{year}{2003}).

\bibitem[{\citenamefont{Brey et~al.}(2006)\citenamefont{Brey, Calder\'on,
  Das~Sarma, and Guinea}}]{Brey2006}
\bibinfo{author}{\bibfnamefont{L.}~\bibnamefont{Brey}},
  \bibinfo{author}{\bibfnamefont{M.~J.} \bibnamefont{Calder\'on}},
  \bibinfo{author}{\bibfnamefont{S.}~\bibnamefont{Das~Sarma}},
  \bibnamefont{and} \bibinfo{author}{\bibfnamefont{F.}~\bibnamefont{Guinea}},
  \bibinfo{journal}{Phys. Rev. B} \textbf{\bibinfo{volume}{74}},
  \bibinfo{pages}{094429} (\bibinfo{year}{2006}).
  
  \bibitem{Majumdar2009}
 P. Sanyal and P. Majumdar, Phys. Rev. B {\bf 80}, 054411 (2009); V. N. Singh and
 P. Majumdar, Europhys. Lett. {\bf 94}, 47004 (2011).

\bibitem[{\citenamefont{Erten et~al.}(2011{\natexlab{a}})\citenamefont{Erten,
  Meetei, Mukherjee, Randeria, Trivedi, and Woodward}}]{Erten_SFMO_PRL2011}
\bibinfo{author}{\bibfnamefont{O.}~\bibnamefont{Erten}},
  \bibinfo{author}{\bibfnamefont{O.~N.} \bibnamefont{Meetei}},
  \bibinfo{author}{\bibfnamefont{A.}~\bibnamefont{Mukherjee}},
  \bibinfo{author}{\bibfnamefont{M.}~\bibnamefont{Randeria}},
  \bibinfo{author}{\bibfnamefont{N.}~\bibnamefont{Trivedi}}, \bibnamefont{and}
  \bibinfo{author}{\bibfnamefont{P.}~\bibnamefont{Woodward}},
  \bibinfo{journal}{Phys. Rev. Lett.} \textbf{\bibinfo{volume}{107}},
  \bibinfo{pages}{257201} (\bibinfo{year}{2011}{\natexlab{a}}).

\bibitem[{\citenamefont{Das et~al.}(2011)\citenamefont{Das, Sanyal,
  Saha-Dasgupta, and Sarma}}]{Das_PRB2011}
\bibinfo{author}{\bibfnamefont{H.}~\bibnamefont{Das}},
  \bibinfo{author}{\bibfnamefont{P.}~\bibnamefont{Sanyal}},
  \bibinfo{author}{\bibfnamefont{T.}~\bibnamefont{Saha-Dasgupta}},
  \bibnamefont{and} \bibinfo{author}{\bibfnamefont{D.~D.} \bibnamefont{Sarma}},
  \bibinfo{journal}{Phys. Rev. B} \textbf{\bibinfo{volume}{83}},
  \bibinfo{pages}{104418} (\bibinfo{year}{2011}).

\bibitem[{\citenamefont{Aharen et~al.}(2010)\citenamefont{Aharen, Greedan,
  Bridges, Aczel, Rodriguez, MacDougall, Luke, Imai, Michaelis, Kroeker
  et~al.}}]{Aharen_PRB2010}
\bibinfo{author}{\bibfnamefont{T.}~\bibnamefont{Aharen}},
  \bibinfo{author}{\bibfnamefont{J.~E.} \bibnamefont{Greedan}},
  \bibinfo{author}{\bibfnamefont{C.~A.} \bibnamefont{Bridges}},
  \bibinfo{author}{\bibfnamefont{A.~A.} \bibnamefont{Aczel}},
  \bibinfo{author}{\bibfnamefont{J.}~\bibnamefont{Rodriguez}},
  \bibinfo{author}{\bibfnamefont{G.}~\bibnamefont{MacDougall}},
  \bibinfo{author}{\bibfnamefont{G.~M.} \bibnamefont{Luke}},
  \bibinfo{author}{\bibfnamefont{T.}~\bibnamefont{Imai}},
  \bibinfo{author}{\bibfnamefont{V.~K.} \bibnamefont{Michaelis}},
  \bibinfo{author}{\bibfnamefont{S.}~\bibnamefont{Kroeker}},
  \bibnamefont{et~al.}, \bibinfo{journal}{Phys. Rev. B}
  \textbf{\bibinfo{volume}{81}}, \bibinfo{pages}{224409}
  (\bibinfo{year}{2010}).

\bibitem[{\citenamefont{Carlo et~al.}(2011)\citenamefont{Carlo, Clancy, Aharen,
  Yamani, Ruff, Wagman, Van~Gastel, Noad, Granroth, Greedan
  et~al.}}]{BYMO_Neutron_PRB2011}
\bibinfo{author}{\bibfnamefont{J.~P.} \bibnamefont{Carlo}},
  \bibinfo{author}{\bibfnamefont{J.~P.} \bibnamefont{Clancy}},
  \bibinfo{author}{\bibfnamefont{T.}~\bibnamefont{Aharen}},
  \bibinfo{author}{\bibfnamefont{Z.}~\bibnamefont{Yamani}},
  \bibinfo{author}{\bibfnamefont{J.~P.~C.} \bibnamefont{Ruff}},
  \bibinfo{author}{\bibfnamefont{J.~J.} \bibnamefont{Wagman}},
  \bibinfo{author}{\bibfnamefont{G.~J.} \bibnamefont{Van~Gastel}},
  \bibinfo{author}{\bibfnamefont{H.~M.~L.} \bibnamefont{Noad}},
  \bibinfo{author}{\bibfnamefont{G.~E.} \bibnamefont{Granroth}},
  \bibinfo{author}{\bibfnamefont{J.~E.} \bibnamefont{Greedan}},
  \bibnamefont{et~al.}, \bibinfo{journal}{Phys. Rev. B}
  \textbf{\bibinfo{volume}{84}}, \bibinfo{pages}{100404}
  (\bibinfo{year}{2011}).

\bibitem[{\citenamefont{Chen et~al.}(2010)\citenamefont{Chen, Pereira, and
  Balents}}]{Chen2010}
\bibinfo{author}{\bibfnamefont{G.}~\bibnamefont{Chen}},
  \bibinfo{author}{\bibfnamefont{R.}~\bibnamefont{Pereira}}, \bibnamefont{and}
  \bibinfo{author}{\bibfnamefont{L.}~\bibnamefont{Balents}},
  \bibinfo{journal}{Phys. Rev. B} \textbf{\bibinfo{volume}{82}},
  \bibinfo{pages}{174440} (\bibinfo{year}{2010}).

\bibitem[{\citenamefont{Dodds et~al.}(2011)\citenamefont{Dodds, Choy, and
  Kim}}]{DoddsPRB2011}
\bibinfo{author}{\bibfnamefont{T.}~\bibnamefont{Dodds}},
  \bibinfo{author}{\bibfnamefont{T.-P.} \bibnamefont{Choy}}, \bibnamefont{and}
  \bibinfo{author}{\bibfnamefont{Y.~B.} \bibnamefont{Kim}},
  \bibinfo{journal}{Phys. Rev. B} \textbf{\bibinfo{volume}{84}},
  \bibinfo{pages}{104439} (\bibinfo{year}{2011}).

\bibitem[{\citenamefont{Chen and Balents}(2011)}]{Chen2011}
\bibinfo{author}{\bibfnamefont{G.}~\bibnamefont{Chen}} \bibnamefont{and}
  \bibinfo{author}{\bibfnamefont{L.}~\bibnamefont{Balents}},
  \bibinfo{journal}{Phys. Rev. B} \textbf{\bibinfo{volume}{84}},
  \bibinfo{pages}{094420} (\bibinfo{year}{2011}).

\bibitem[{\citenamefont{\ifmmode \check{Z}\else
  \v{Z}\fi{}uti\ifmmode~\acute{c}\else \'{c}\fi{}
  et~al.}(2004)\citenamefont{\ifmmode \check{Z}\else
  \v{Z}\fi{}uti\ifmmode~\acute{c}\else \'{c}\fi{}, Fabian, and
  Das~Sarma}}]{Zutic2004}
\bibinfo{author}{\bibfnamefont{I.}~\bibnamefont{\ifmmode \check{Z}\else
  \v{Z}\fi{}uti\ifmmode~\acute{c}\else \'{c}\fi{}}},
  \bibinfo{author}{\bibfnamefont{J.}~\bibnamefont{Fabian}}, \bibnamefont{and}
  \bibinfo{author}{\bibfnamefont{S.}~\bibnamefont{Das~Sarma}},
  \bibinfo{journal}{Rev. Mod. Phys.} \textbf{\bibinfo{volume}{76}},
  \bibinfo{pages}{323} (\bibinfo{year}{2004}).

\bibitem[{\citenamefont{Alff}(2007)}]{Alff_Springer2007}
\bibinfo{author}{\bibfnamefont{L.}~\bibnamefont{Alff}}, in
  \emph{\bibinfo{booktitle}{Electron Correlation in New Materials and
  Nanosystems}}, edited by
  \bibinfo{editor}{\bibfnamefont{K.}~\bibnamefont{Scharnberg}}
  \bibnamefont{and} \bibinfo{editor}{\bibfnamefont{S.}~\bibnamefont{Kruchinin}}
  (\bibinfo{publisher}{Springer Netherlands}, \bibinfo{year}{2007}), vol.
  \bibinfo{volume}{241} of \emph{\bibinfo{series}{NATO Science Series}}, pp.
  \bibinfo{pages}{393--400}.

\bibitem[{\citenamefont{Plumb et~al.}(2013)\citenamefont{Plumb, Cook, Clancy,
  Kolesnikov, Jeon, Noh, Paramekanti, and Kim}}]{Plumb_PRB2013}
\bibinfo{author}{\bibfnamefont{K.~W.} \bibnamefont{Plumb}},
  \bibinfo{author}{\bibfnamefont{A.~M.} \bibnamefont{Cook}},
  \bibinfo{author}{\bibfnamefont{J.~P.} \bibnamefont{Clancy}},
  \bibinfo{author}{\bibfnamefont{A.~I.} \bibnamefont{Kolesnikov}},
  \bibinfo{author}{\bibfnamefont{B.~C.} \bibnamefont{Jeon}},
  \bibinfo{author}{\bibfnamefont{T.~W.} \bibnamefont{Noh}},
  \bibinfo{author}{\bibfnamefont{A.}~\bibnamefont{Paramekanti}},
  \bibnamefont{and} \bibinfo{author}{\bibfnamefont{Y.-J.} \bibnamefont{Kim}},
  \bibinfo{journal}{Phys. Rev. B} \textbf{\bibinfo{volume}{87}},
  \bibinfo{pages}{184412} (\bibinfo{year}{2013}).

\bibitem[{\citenamefont{{Witczak-Krempa}
  et~al.}(2013)\citenamefont{{Witczak-Krempa}, {Chen}, {Kim}, and
  {Balents}}}]{Iridate_Review}
\bibinfo{author}{\bibfnamefont{W.}~\bibnamefont{{Witczak-Krempa}}},
  \bibinfo{author}{\bibfnamefont{G.}~\bibnamefont{{Chen}}},
  \bibinfo{author}{\bibfnamefont{Y.~B.} \bibnamefont{{Kim}}}, \bibnamefont{and}
  \bibinfo{author}{\bibfnamefont{L.}~\bibnamefont{{Balents}}},
  \bibinfo{journal}{ArXiv e-prints}  (\bibinfo{year}{2013}),
  \eprint{1305.2193}.

\bibitem[{\citenamefont{Levin and Stern}(2009)}]{Levin_PRL2009}
\bibinfo{author}{\bibfnamefont{M.}~\bibnamefont{Levin}} \bibnamefont{and}
  \bibinfo{author}{\bibfnamefont{A.}~\bibnamefont{Stern}},
  \bibinfo{journal}{Phys. Rev. Lett.} \textbf{\bibinfo{volume}{103}},
  \bibinfo{pages}{196803} (\bibinfo{year}{2009}).

\bibitem[{\citenamefont{Pesin and Balents}(2010)}]{Pesin_NPhys2010}
\bibinfo{author}{\bibfnamefont{D.}~\bibnamefont{Pesin}} \bibnamefont{and}
  \bibinfo{author}{\bibfnamefont{L.}~\bibnamefont{Balents}},
  \bibinfo{journal}{Nature Physics} \textbf{\bibinfo{volume}{6}},
  \bibinfo{pages}{376} (\bibinfo{year}{2010}).

\bibitem[{\citenamefont{Maciejko et~al.}(2010)\citenamefont{Maciejko, Qi,
  Karch, and Zhang}}]{Maciejko_PRL2010}
\bibinfo{author}{\bibfnamefont{J.}~\bibnamefont{Maciejko}},
  \bibinfo{author}{\bibfnamefont{X.-L.} \bibnamefont{Qi}},
  \bibinfo{author}{\bibfnamefont{A.}~\bibnamefont{Karch}}, \bibnamefont{and}
  \bibinfo{author}{\bibfnamefont{S.-C.} \bibnamefont{Zhang}},
  \bibinfo{journal}{Phys. Rev. Lett.} \textbf{\bibinfo{volume}{105}},
  \bibinfo{pages}{246809} (\bibinfo{year}{2010}).

\bibitem[{\citenamefont{Swingle et~al.}(2011)\citenamefont{Swingle, Barkeshli,
  McGreevy, and Senthil}}]{Swingle_PRB2011}
\bibinfo{author}{\bibfnamefont{B.}~\bibnamefont{Swingle}},
  \bibinfo{author}{\bibfnamefont{M.}~\bibnamefont{Barkeshli}},
  \bibinfo{author}{\bibfnamefont{J.}~\bibnamefont{McGreevy}}, \bibnamefont{and}
  \bibinfo{author}{\bibfnamefont{T.}~\bibnamefont{Senthil}},
  \bibinfo{journal}{Phys. Rev. B} \textbf{\bibinfo{volume}{83}},
  \bibinfo{pages}{195139} (\bibinfo{year}{2011}).

\bibitem[{\citenamefont{Wan et~al.}(2011)\citenamefont{Wan, Turner, Vishwanath,
  and Savrasov}}]{Wan_PRB2011}
\bibinfo{author}{\bibfnamefont{X.}~\bibnamefont{Wan}},
  \bibinfo{author}{\bibfnamefont{A.~M.} \bibnamefont{Turner}},
  \bibinfo{author}{\bibfnamefont{A.}~\bibnamefont{Vishwanath}},
  \bibnamefont{and} \bibinfo{author}{\bibfnamefont{S.~Y.}
  \bibnamefont{Savrasov}}, \bibinfo{journal}{Phys. Rev. B}
  \textbf{\bibinfo{volume}{83}}, \bibinfo{pages}{205101}
  (\bibinfo{year}{2011}).

\bibitem[{\citenamefont{Burkov and Balents}(2011)}]{Burkov_PRL2011}
\bibinfo{author}{\bibfnamefont{A.~A.} \bibnamefont{Burkov}} \bibnamefont{and}
  \bibinfo{author}{\bibfnamefont{L.}~\bibnamefont{Balents}},
  \bibinfo{journal}{Phys. Rev. Lett.} \textbf{\bibinfo{volume}{107}},
  \bibinfo{pages}{127205} (\bibinfo{year}{2011}).

\bibitem[{\citenamefont{Yang et~al.}(2011)\citenamefont{Yang, Lu, and
  Ran}}]{YRan_PRB2011}
\bibinfo{author}{\bibfnamefont{K.-Y.} \bibnamefont{Yang}},
  \bibinfo{author}{\bibfnamefont{Y.-M.} \bibnamefont{Lu}}, \bibnamefont{and}
  \bibinfo{author}{\bibfnamefont{Y.}~\bibnamefont{Ran}},
  \bibinfo{journal}{Phys. Rev. B} \textbf{\bibinfo{volume}{84}},
  \bibinfo{pages}{075129} (\bibinfo{year}{2011}).

\bibitem[{\citenamefont{Witczak-Krempa and Kim}(2012)}]{Witczak2011}
\bibinfo{author}{\bibfnamefont{W.}~\bibnamefont{Witczak-Krempa}}
  \bibnamefont{and} \bibinfo{author}{\bibfnamefont{Y.~B.} \bibnamefont{Kim}},
  \bibinfo{journal}{Phys. Rev. B} \textbf{\bibinfo{volume}{85}},
  \bibinfo{pages}{045124} (\bibinfo{year}{2012}).

\bibitem[{\citenamefont{Sleight and Weiher}(1972)}]{Sleight1972}
\bibinfo{author}{\bibfnamefont{A.~W.} \bibnamefont{Sleight}} \bibnamefont{and}
  \bibinfo{author}{\bibfnamefont{J.~F.} \bibnamefont{Weiher}},
  \bibinfo{journal}{J. Phys. Chem. Solids} \textbf{\bibinfo{volume}{33}},
  \bibinfo{pages}{679} (\bibinfo{year}{1972}).

\bibitem[{\citenamefont{Prellier et~al.}(2000)\citenamefont{Prellier,
  Smolyaninova, Biswas, Galley, Greene, Ramesha, and
  Gopalakrishnan}}]{Prellier2000}
\bibinfo{author}{\bibfnamefont{W.}~\bibnamefont{Prellier}},
  \bibinfo{author}{\bibfnamefont{V.}~\bibnamefont{Smolyaninova}},
  \bibinfo{author}{\bibfnamefont{A.}~\bibnamefont{Biswas}},
  \bibinfo{author}{\bibfnamefont{C.}~\bibnamefont{Galley}},
  \bibinfo{author}{\bibfnamefont{R.~L.} \bibnamefont{Greene}},
  \bibinfo{author}{\bibfnamefont{K.}~\bibnamefont{Ramesha}}, \bibnamefont{and}
  \bibinfo{author}{\bibfnamefont{J.}~\bibnamefont{Gopalakrishnan}},
  \bibinfo{journal}{J. Phys.: Condens. Matter} \textbf{\bibinfo{volume}{12}},
  \bibinfo{pages}{965} (\bibinfo{year}{2000}).

\bibitem[{\citenamefont{Jeon et~al.}(2010)\citenamefont{Jeon, Kim, Moon, Choi,
  Jeong, Lee, Yu, Won, Jung, Hur et~al.}}]{Jeon2010}
\bibinfo{author}{\bibfnamefont{B.~C.} \bibnamefont{Jeon}},
  \bibinfo{author}{\bibfnamefont{C.~H.} \bibnamefont{Kim}},
  \bibinfo{author}{\bibfnamefont{S.~J.} \bibnamefont{Moon}},
  \bibinfo{author}{\bibfnamefont{W.~S.} \bibnamefont{Choi}},
  \bibinfo{author}{\bibfnamefont{H.}~\bibnamefont{Jeong}},
  \bibinfo{author}{\bibfnamefont{Y.~S.} \bibnamefont{Lee}},
  \bibinfo{author}{\bibfnamefont{J.}~\bibnamefont{Yu}},
  \bibinfo{author}{\bibfnamefont{C.~J.} \bibnamefont{Won}},
  \bibinfo{author}{\bibfnamefont{J.~H.} \bibnamefont{Jung}},
  \bibinfo{author}{\bibfnamefont{N.}~\bibnamefont{Hur}}, \bibnamefont{et~al.},
  \bibinfo{journal}{J. Phys.: Condens. Matter} \textbf{\bibinfo{volume}{22}},
  \bibinfo{pages}{345602} (\bibinfo{year}{2010}).

\bibitem[{\citenamefont{Teresa et~al.}(2007)\citenamefont{Teresa, Michalik,
  Blasco, Algarabel, Ibarra, Kapusta, and Zeitler}}]{Teresa2007}
\bibinfo{author}{\bibfnamefont{J.~M.~D.} \bibnamefont{Teresa}},
  \bibinfo{author}{\bibfnamefont{J.~M.} \bibnamefont{Michalik}},
  \bibinfo{author}{\bibfnamefont{J.}~\bibnamefont{Blasco}},
  \bibinfo{author}{\bibfnamefont{P.~A.} \bibnamefont{Algarabel}},
  \bibinfo{author}{\bibfnamefont{M.~R.} \bibnamefont{Ibarra}},
  \bibinfo{author}{\bibfnamefont{C.}~\bibnamefont{Kapusta}}, \bibnamefont{and}
  \bibinfo{author}{\bibfnamefont{U.}~\bibnamefont{Zeitler}},
  \bibinfo{journal}{Applied Physics Letters} \textbf{\bibinfo{volume}{90}},
  \bibinfo{eid}{252514} (\bibinfo{year}{2007}).

\bibitem[{\citenamefont{Azimonte et~al.}(2007)\citenamefont{Azimonte, Cezar,
  Granado, Huang, Lynn, Campoy, Gopalakrishnan, and Ramesha}}]{Azimonte2007}
\bibinfo{author}{\bibfnamefont{C.}~\bibnamefont{Azimonte}},
  \bibinfo{author}{\bibfnamefont{J.~C.} \bibnamefont{Cezar}},
  \bibinfo{author}{\bibfnamefont{E.}~\bibnamefont{Granado}},
  \bibinfo{author}{\bibfnamefont{Q.}~\bibnamefont{Huang}},
  \bibinfo{author}{\bibfnamefont{J.~W.} \bibnamefont{Lynn}},
  \bibinfo{author}{\bibfnamefont{J.~C.~P.} \bibnamefont{Campoy}},
  \bibinfo{author}{\bibfnamefont{J.}~\bibnamefont{Gopalakrishnan}},
  \bibnamefont{and} \bibinfo{author}{\bibfnamefont{K.}~\bibnamefont{Ramesha}},
  \bibinfo{journal}{Phys. Rev. Lett.} \textbf{\bibinfo{volume}{98}},
  \bibinfo{pages}{017204} (\bibinfo{year}{2007}).

\bibitem[{\citenamefont{Winkler et~al.}(2009)\citenamefont{Winkler, Narayanan,
  Mikhailova, Bramnik, Ehrenberg, Fuess, Vaitheeswaran, Kanchana, Wilhelm,
  Rogalev et~al.}}]{Winkler_NJP2009}
\bibinfo{author}{\bibfnamefont{A.}~\bibnamefont{Winkler}},
  \bibinfo{author}{\bibfnamefont{N.}~\bibnamefont{Narayanan}},
  \bibinfo{author}{\bibfnamefont{D.}~\bibnamefont{Mikhailova}},
  \bibinfo{author}{\bibfnamefont{K.~G.} \bibnamefont{Bramnik}},
  \bibinfo{author}{\bibfnamefont{H.}~\bibnamefont{Ehrenberg}},
  \bibinfo{author}{\bibfnamefont{H.}~\bibnamefont{Fuess}},
  \bibinfo{author}{\bibfnamefont{G.}~\bibnamefont{Vaitheeswaran}},
  \bibinfo{author}{\bibfnamefont{V.}~\bibnamefont{Kanchana}},
  \bibinfo{author}{\bibfnamefont{F.}~\bibnamefont{Wilhelm}},
  \bibinfo{author}{\bibfnamefont{A.}~\bibnamefont{Rogalev}},
  \bibnamefont{et~al.}, \bibinfo{journal}{New Journal of Physics}
  \textbf{\bibinfo{volume}{11}}, \bibinfo{pages}{073047}
  (\bibinfo{year}{2009}).

\bibitem[{\citenamefont{Burkov et~al.}(2011)\citenamefont{Burkov, Hook, and
  Balents}}]{Burkov_PRB2011}
\bibinfo{author}{\bibfnamefont{A.~A.} \bibnamefont{Burkov}},
  \bibinfo{author}{\bibfnamefont{M.~D.} \bibnamefont{Hook}}, \bibnamefont{and}
  \bibinfo{author}{\bibfnamefont{L.}~\bibnamefont{Balents}},
  \bibinfo{journal}{Phys. Rev. B} \textbf{\bibinfo{volume}{84}},
  \bibinfo{pages}{235126} (\bibinfo{year}{2011}).

\bibitem[{\citenamefont{Hal\'asz and Balents}(2012)}]{Halasz_PRB2012}
\bibinfo{author}{\bibfnamefont{G.~B.} \bibnamefont{Hal\'asz}} \bibnamefont{and}
  \bibinfo{author}{\bibfnamefont{L.}~\bibnamefont{Balents}},
  \bibinfo{journal}{Phys. Rev. B} \textbf{\bibinfo{volume}{85}},
  \bibinfo{pages}{035103} (\bibinfo{year}{2012}).

\bibitem[{\citenamefont{Erten et~al.}(2011{\natexlab{b}})\citenamefont{Erten,
  Meetei, Mukherjee, Randeria, Trivedi, and Woodward}}]{Erten2011}
\bibinfo{author}{\bibfnamefont{O.}~\bibnamefont{Erten}},
  \bibinfo{author}{\bibfnamefont{O.~N.} \bibnamefont{Meetei}},
  \bibinfo{author}{\bibfnamefont{A.}~\bibnamefont{Mukherjee}},
  \bibinfo{author}{\bibfnamefont{M.}~\bibnamefont{Randeria}},
  \bibinfo{author}{\bibfnamefont{N.}~\bibnamefont{Trivedi}}, \bibnamefont{and}
  \bibinfo{author}{\bibfnamefont{P.}~\bibnamefont{Woodward}},
  \bibinfo{journal}{Phys. Rev. Lett.} \textbf{\bibinfo{volume}{107}},
  \bibinfo{pages}{257201} (\bibinfo{year}{2011}{\natexlab{b}}).

\bibitem[{\citenamefont{Erten et~al.}(2013)\citenamefont{Erten, Meetei,
  Mukherjee, Randeria, Trivedi, and Woodward}}]{ErtenPRB2013}
\bibinfo{author}{\bibfnamefont{O.}~\bibnamefont{Erten}},
  \bibinfo{author}{\bibfnamefont{O.~N.} \bibnamefont{Meetei}},
  \bibinfo{author}{\bibfnamefont{A.}~\bibnamefont{Mukherjee}},
  \bibinfo{author}{\bibfnamefont{M.}~\bibnamefont{Randeria}},
  \bibinfo{author}{\bibfnamefont{N.}~\bibnamefont{Trivedi}}, \bibnamefont{and}
  \bibinfo{author}{\bibfnamefont{P.}~\bibnamefont{Woodward}},
  \bibinfo{journal}{Phys. Rev. B} \textbf{\bibinfo{volume}{87}},
  \bibinfo{pages}{165105} (\bibinfo{year}{2013}).

\bibitem[{\citenamefont{Meetei et~al.}(2013)\citenamefont{Meetei, Erten,
  Mukherjee, Randeria, Trivedi, and Woodward}}]{MeeteiPRB2013}
\bibinfo{author}{\bibfnamefont{O.~N.} \bibnamefont{Meetei}},
  \bibinfo{author}{\bibfnamefont{O.}~\bibnamefont{Erten}},
  \bibinfo{author}{\bibfnamefont{A.}~\bibnamefont{Mukherjee}},
  \bibinfo{author}{\bibfnamefont{M.}~\bibnamefont{Randeria}},
  \bibinfo{author}{\bibfnamefont{N.}~\bibnamefont{Trivedi}}, \bibnamefont{and}
  \bibinfo{author}{\bibfnamefont{P.}~\bibnamefont{Woodward}},
  \bibinfo{journal}{Phys. Rev. B} \textbf{\bibinfo{volume}{87}},
  \bibinfo{pages}{165104} (\bibinfo{year}{2013}).

\bibitem[{\citenamefont{Fazekas}(1999)}]{Fazekas}
\bibinfo{author}{\bibfnamefont{P.}~\bibnamefont{Fazekas}},
  \emph{\bibinfo{title}{Lectures Notes on Electron Correlation and Magnetism}}
  (\bibinfo{publisher}{World Scientific}, \bibinfo{year}{1999}).

\bibitem[{\citenamefont{Gopalakrishnan
  et~al.}(2000)\citenamefont{Gopalakrishnan, Chattopadhyay, Ogale, Venkatesan,
  Greene, Millis, Ramesha, Hannoyer, and Marest}}]{Gopalakrishnan_PRB2000}
\bibinfo{author}{\bibfnamefont{J.}~\bibnamefont{Gopalakrishnan}},
  \bibinfo{author}{\bibfnamefont{A.}~\bibnamefont{Chattopadhyay}},
  \bibinfo{author}{\bibfnamefont{S.~B.} \bibnamefont{Ogale}},
  \bibinfo{author}{\bibfnamefont{T.}~\bibnamefont{Venkatesan}},
  \bibinfo{author}{\bibfnamefont{R.~L.} \bibnamefont{Greene}},
  \bibinfo{author}{\bibfnamefont{A.~J.} \bibnamefont{Millis}},
  \bibinfo{author}{\bibfnamefont{K.}~\bibnamefont{Ramesha}},
  \bibinfo{author}{\bibfnamefont{B.}~\bibnamefont{Hannoyer}}, \bibnamefont{and}
  \bibinfo{author}{\bibfnamefont{G.}~\bibnamefont{Marest}},
  \bibinfo{journal}{Phys. Rev. B} \textbf{\bibinfo{volume}{62}},
  \bibinfo{pages}{9538} (\bibinfo{year}{2000}).

\bibitem[{\citenamefont{{Chen} et~al.}(2013)\citenamefont{{Chen}, {Bergman},
  and {Burkov}}}]{Burkov_WeylAHE}
\bibinfo{author}{\bibfnamefont{Y.}~\bibnamefont{{Chen}}},
  \bibinfo{author}{\bibfnamefont{D.~L.} \bibnamefont{{Bergman}}},
  \bibnamefont{and} \bibinfo{author}{\bibfnamefont{A.~A.}
  \bibnamefont{{Burkov}}}, \bibinfo{journal}{ArXiv e-prints}
  (\bibinfo{year}{2013}), \eprint{1305.0183}.

\bibitem[{\citenamefont{Sundaram and Niu}(1999)}]{Niu_PRB1999}
\bibinfo{author}{\bibfnamefont{G.}~\bibnamefont{Sundaram}} \bibnamefont{and}
  \bibinfo{author}{\bibfnamefont{Q.}~\bibnamefont{Niu}},
  \bibinfo{journal}{Phys. Rev. B} \textbf{\bibinfo{volume}{59}},
  \bibinfo{pages}{14915} (\bibinfo{year}{1999}).

\bibitem[{\citenamefont{Nagaosa et~al.}(2010)\citenamefont{Nagaosa, Sinova,
  Onoda, MacDonald, and Ong}}]{RMP_AHE}
\bibinfo{author}{\bibfnamefont{N.}~\bibnamefont{Nagaosa}},
  \bibinfo{author}{\bibfnamefont{J.}~\bibnamefont{Sinova}},
  \bibinfo{author}{\bibfnamefont{S.}~\bibnamefont{Onoda}},
  \bibinfo{author}{\bibfnamefont{A.~H.} \bibnamefont{MacDonald}},
  \bibnamefont{and} \bibinfo{author}{\bibfnamefont{N.~P.} \bibnamefont{Ong}},
  \bibinfo{journal}{Rev. Mod. Phys.} \textbf{\bibinfo{volume}{82}},
  \bibinfo{pages}{1539} (\bibinfo{year}{2010}).

\bibitem[{\citenamefont{Fukui et~al.}(2005)\citenamefont{Fukui, Hatsugai, and
  Suzuki}}]{Hatsugai_JPSJ2005}
\bibinfo{author}{\bibfnamefont{T.}~\bibnamefont{Fukui}},
  \bibinfo{author}{\bibfnamefont{Y.}~\bibnamefont{Hatsugai}}, \bibnamefont{and}
  \bibinfo{author}{\bibfnamefont{H.}~\bibnamefont{Suzuki}},
  \bibinfo{journal}{Journal of the Physical Society of Japan}
  \textbf{\bibinfo{volume}{74}}, \bibinfo{pages}{1674} (\bibinfo{year}{2005}).

\end{thebibliography}
\end{document}